\documentclass[a4paper,aps,preprint,pra]{revtex4-1}
\usepackage[english]{babel}
\usepackage{graphicx}
\usepackage{latexsym}
\usepackage{amsmath}
\usepackage{amsfonts}
\usepackage{amssymb}

\newcommand{\E}{\mathrm{e}}

\begin{document}

\title{Analogue of the quantum Hall effect for neutral particles with magnetic dipole moment}

\author{L. R. Ribeiro}\email{passos@df.ufcg.edu.br}
\affiliation{{Unidade Acadêmica de Ciência e Tecnologia Ambiental, Universidade Federal de Campina Grande, 58840-000, Pombal, PB, Brasil.}}

\author{E. Passos}
\email{passos@df.ufcg.edu.br}
\affiliation{{ Departamento de F´ısica, Universidade Federal de Campina Grande
Caixa Postal 10071, 58429-900 Campina Grande, Para´ıba, Brazil}}
\affiliation{Instituto de F\' isica, Universidade Federal do Rio de Janeiro, Caixa Postal 21945, Rio de Janeiro, 
Rio de Janeiro, Brazil.}

\author{C. Furtado}
\email{furtado@fisica.ufpb.br}
\affiliation{{ Departamento de
F\'{\i}sica, Universidade Federal da Para\'\i ba, Caixa Postal 5008, 58051-970,
Jo\~ao Pessoa, PB, Brasil}}

\author{S. Sergeenkov}
\email{sergei@fisica.ufpb.br}
\affiliation{{ Departamento de
F\'{\i}sica, Universidade Federal da Para\'\i ba, Caixa Postal 5008, 58051-970,
Jo\~ao Pessoa, PB, Brasil}}

\begin{abstract}
In this paper we investigate a possibility for the existence of an analog of the Quantum Hall Effect for neutral particles with a permanent magnetic moment $\mu$ in the presence of crossed inhomogeneous magnetic and electric fields. We predict the appearance of Hall conductivity $\sigma_H=(e^2/h)\nu (\mu)$ with the Landau filling factor $\nu (\mu)\propto \mu ^2$. The estimates of the model parameters suggest quite an optimistic possibility to experimentally verify this prediction in optically trapped clouds of atomic BEC. 
\end{abstract}

\maketitle
\section{Introduction}
  
Recall that in strong enough magnetic fields, the charged particles give rise to the Hall effect due to the $\vec{E} \times \vec{B}$ drift in the crossed electric $\vec{E}$ and magnetic $\vec{B}$ fields. According to Wolfgang Ketterle \cite{Ketterle}, the neutral atoms can be used as quantum simulators for electrons in a strong magnetic field. This is accomplished by creating synthetic magnetic fields with the help of laser beams. Recently, Monika Aidelsburger and colleagues \cite{Monika} have studied the dynamics of ultracold atoms in crossed electric and magnetic fields and observed the manifestation of the Quantum Hall Effect (QHE).

It is important to mention that nowadays QHE \cite{klitz, tsui,laughlin, prange, comtet, dunne} is studied in the context of many different phenomena including the non-commutative geometry \cite{nct,dayi}, atomic systems  \cite{paredes1,paredes2}, curved spaces \cite{hiperbolic}, droplets \cite{droplet}, etc. 
In addition, there has been an immense progress in better understanding of the quantum properties of neutral particles undergoing the Bose-Einstein condensation (BEC) \cite{BEC,sergei}. 

On the other hand, Ericsson and Sj\"{o}qvist \cite{ericsson} investigated the quantum dynamics of a neutral particle with a permanent magnetic moment interacting with an electric field configuration via the Aharonov-Casher type  coupling \cite{aha-casher}. They have demonstrated that an analog of the Landau quantization takes place only when a specific field-dipole configuration is realized. Following the same approach of Ericsson and Sj\"{o}qvist, the  analog of Landau quantization was predicted to occur for a permanent \cite{pla1}  and induced \cite{pla2} electric dipoles and for an electric quadrupole system \cite{quadrupole}. Several authors have investigated the trapping of neutral particles inside the external fields  \cite{paul,migdal,schmiedmayer,comptom,lin,jacob} as well as the appearance of degenerate Landau levels for tripod-type cold atoms with Aharonov-Casher coupling \cite{bruno, bruno2, bruno3}. An important study on Landau-like quantization in neutral atoms with magnetic dipole systems using spin-orbit coupling was recently presented by Banerjee {\it et al.}\cite{banerjee}.

The main purpose of this paper is to investigate nontrivial consequences of the Aharonov-Casher coupling in neutral boson particles with dipole magnetic moments under the influence of crossed gradient-containing external magnetic and electric fields.  As a result, an unexpected manifestation of the Compton physics has surfaced within the scope of the atomic BEC leading to appearance of a new QHE type behavior. A possibility to observe the predicted effects is discussed.

\section{Landau levels for neutral particles with magnetic dipole moments}

In this Section we discuss the structure of the Landau levels spectrum for a neutral (chargeless) particle with a permanent magnetic moment $\vec{\mu}=\mu \vec{z}$, where $\vec{z}=(0,0,1)$, in the presence of crossed inhomogeneous electric $\vec{E}$ and magnetic $\vec{B}$ fields.
The appropriate form of the Hamiltonian for treating this problem reads
\begin{equation}
	H=\frac{1}{2m}\left(\vec{p}-\frac{\vec{\mu}\times\vec{E}}{c^2}\right)^2-\frac{\mu\hbar}{2mc^2}\nabla\cdot\vec{E}+ \vec{\mu}\cdot\vec{B}\;,
	\label{eq:hc1}
\end{equation}
In what follows, we assume these explicit forms for the inhomogeneous external fields
\begin{eqnarray}
\vec{E}(x)=(\alpha x,0,0)  \text{ and  } \vec{B}(x)=(0,0, \beta x).
\end{eqnarray}   \label{eq:hc1.2}
Now, we may rewrite the Hamiltonian (\ref{eq:hc1}) as follows
\begin{equation}
	H=\frac{p_x^2}{2m}+\frac12m\omega^2\left[x-\left(p_y\frac{\ell^2}{\hbar}-\frac{\mu \beta}{m\omega^2}\right)\right]^2-\frac{\hbar \omega}{2}+\mu \beta\left(p_y\frac{\ell^2}{\hbar}-\frac{\mu \beta}{2m\omega^2}\right)\,
	\label{eq:hc6}
\end{equation}
where $\omega=\frac{\mu \alpha}{mc^2}$ is the cyclotron frequency and $\ell=\sqrt{\frac{\hbar c^2}{\mu\alpha}}$ is the fundamental length (analog of the magnetic length for conventional Landau levels).  Due to the Field-dipole configuration, the problem becomes effectively two-dimensional, for this reason we ignore the term of the linear momentum $p_{z}$ in the $z$-direction  in Eq.(\ref{eq:hc6}) \cite{ericsson}.

Using the Ansatz $\Psi(x,y)=\E^{iky}\chi(x)$ for the wave function and defining $X_k=\left(k\frac{\ell^2}{\hbar}-\frac{\mu \beta}{m\omega^2}\right)$ as the center of harmonic oscillator, the reduced Hamiltonian acting on $\chi(x)$ reads
\begin{equation}
	H=\frac{p_x^2}{2m}+\frac12m\omega^2\left(x-X_k\right)^2-\frac{\hbar \omega}{2}+\mu \beta X_k+\frac12mv_\textrm{D}^2\;
	\label{eq:hc7}
\end{equation}
where $v_{D}=c^2\frac{|\vec{E}\times\vec{B}|}{|\vec{E}|^2}=\frac{\beta c^2}{\alpha}$ is the drift velocity. 

In the most interesting case of strong magnetic fields, the lowest level of Landau spectrum (corresponding to $n=0$) becomes decoupled from the excited states (with $n > 0$). In this approximation, the eigenfunctions $\Psi_0(x,y)$ and the corresponding eigenvalues $\mathcal{E}_{0k}$ defined as the solutions of the equation $H\Psi_0(x,y)=\mathcal{E}_{0k}\Psi_0(x,y)$ are given by

\begin{equation}
	\Psi_0(x,y)=\frac{1}{\sqrt{\ell\sqrt{\pi}}}\E^{-\frac{1}{2\ell^2}(x-X_k)^2}\E^{iky}
	\label{eq:hc9}
\end{equation}

and

\begin{equation}
	\mathcal{E}_{0k}=\hbar |\omega|\left(\frac{1-\sigma}{2}\right)+\mu \beta X_k+\frac{1}{2}mv_\textrm{D}^2
	\label{eq:hc8}
\end{equation}
where $\sigma=\pm$ labels the revolution direction of the cyclotron frequency $\omega=\sigma|\omega|$.

\section{Results and Discussion}

Using the above-obtained solutions, we can calculate the zero-temperature expectation value of the current density for a neutral particle with the constant magnetic moment $\mu$. The result is as follows: 
\begin{equation}
	\langle\vec{J}\rangle=\frac{\mu\varrho}{m\hbar}\langle\Psi_0|\vec{p}-\frac{\mu \vec{A}_\textrm{eff}}{c^2}|\Psi_0\rangle\label{eq:hc10}
\end{equation}
where $\varrho=mN/S$ is a mass density for $N$ neutral particles of mass $m$ over the projected area $S$, and $\vec{A}_\textrm{eff}=\vec{z}\times \vec{E}=(0,\alpha x,0)$ is an effective vector potential.

It can be verified that the $x$-component of the current is zero, $\langle J_x\rangle=\frac{\mu\varrho}{m\hbar}\langle\Psi_0|p_x|\Psi_0\rangle=0$ while the remaining $y$-component is given by 
\begin{eqnarray}
	\langle J_y \rangle &=&\frac{\mu\varrho}{m\hbar}\langle\Psi_0|p_y-\frac{\mu \alpha x}{c^2}|\Psi_0\rangle =\frac{\mu}{\hbar}\varrho v_\textrm{D}\label{eq:hc11}
\end{eqnarray}
Let us demonstrate that $\langle J_y \rangle$ is indeed the \textit{electric} current density. Notice that the factor $\mu/\hbar$ changes the mass density $\varrho=mN/S$ to charge density $\varrho_e=eN/S$. Assuming that $\mu$ is the Bohr magneton $\mu_B=e\hbar/2m$, we obtain $\mu_B\varrho/\hbar=e\varrho/2m=eN/2S=\varrho_e/2$.

Based on the above results, we can find now the expression for the magnetic moment induced conductivity of neutral particles
\begin{equation}
	\left(\begin{array}{c}
	\langle J_x\rangle\\\langle J_y\rangle
	\end{array}\right)=
	\left(\begin{array}{cc}
	\sigma_{xx}&\sigma_{xy}\\
	\sigma_{yx}&\sigma_{yy}
	\end{array}\right)
	\left(\begin{array}{c}
	\langle E_x\rangle\\\langle E_y\rangle
	\end{array}\right)\;
	\label{eq:hc12}
\end{equation}
where $\sigma_{ij}$ is the matrix formed by the conductivity tensor,  $\langle E_x\rangle =\alpha l$  and $\langle E_y\rangle = 0$. Thus, the seeking Hall conductivity $\sigma_{H} \equiv \sigma_{yx}=\langle J_y\rangle/\langle E_x\rangle$ is given by  
\begin{equation}
	\sigma_H=\frac{e^2}{h}\nu (\mu) 
	\label{eq:15}
\end{equation}
where 
\begin{equation}
\nu(\mu)=\frac{\rho_N \lambda_C^2}{2\pi}\left(\frac{\mu }{\mu_B}\right)^2
\end{equation}
is the Landau filling factor with $\rho_N=N/S$ the particle number density and $\lambda_C=\frac{h}{mc}$ the Compton wavelength.

Turning to the discussion of the obtained results, we note that introduced in this paper gradients of crossed inhomogeneous electric and magnetic fields provide an opportunity to test our predictions experimentally by using the optical gradient technique to induce Bloch oscillations in ultracold atomic BEC \cite{Monika}. To be more specific, let us estimate the value of the filling factor $\nu (\mu)$ for atoms of $^{87}Rb$. Assuming that the atomic cloud containing as many as \cite{BEC} $N=10^8$ dipoles with a total magnetic moment $\mu =N\mu_B(Rb)$ (where $\mu_B(Rb)=e\hbar/2m_{Rb}$ is the Bohr magneton for $Rb$ atom with mass $m_{Rb}$ corresponding to the Compton wavelength $\lambda_C \simeq 10^{-17}m$) is trapped within a projected area of $S=100\mu m^2$, Eq.(10) predicts $\nu (\mu)\simeq 1$ which corresponds to the first Chern number deduced from the recently observed QHE in neutral BEC \cite{Monika}.

\section{Conclusion}

In summary, we discussed an analog of the Quantum Hall Effect for a neutral particle with permanent magnetic moment in the presence of crossed inhomogeneous electric and magnetic fields.
The obtained results suggest quite a realistic possibility to experimentally observe this novel effect in cold atoms undergoing the Bose-Einstein condensation scenario. Finally, we claim  that another important point to be considered in an experimental arrangement is that the fields configurations $\vec{E}-\vec{B}$, given by Eq.(2), requires a  charge density inside of the  the region where the neutral particle moves. Although experimentally challenging, such an arrangement could be composed of a two-dimensional toroidal trap and a perpendicular line of charge passing through the
center of the toroid \cite{wood}.

{\bf Acknowledgments.} This work was partially supported by the Con\-se\-lho Na\-cio\-nal de De\-sen\-vol\-vi\-men\-to Cient\'{\i}fico e Tecnol\'{o}gico (CNPq), CAPES,  Projeto Universal/CNPQ and FAPESQ (DCR-PB).

\end{document}